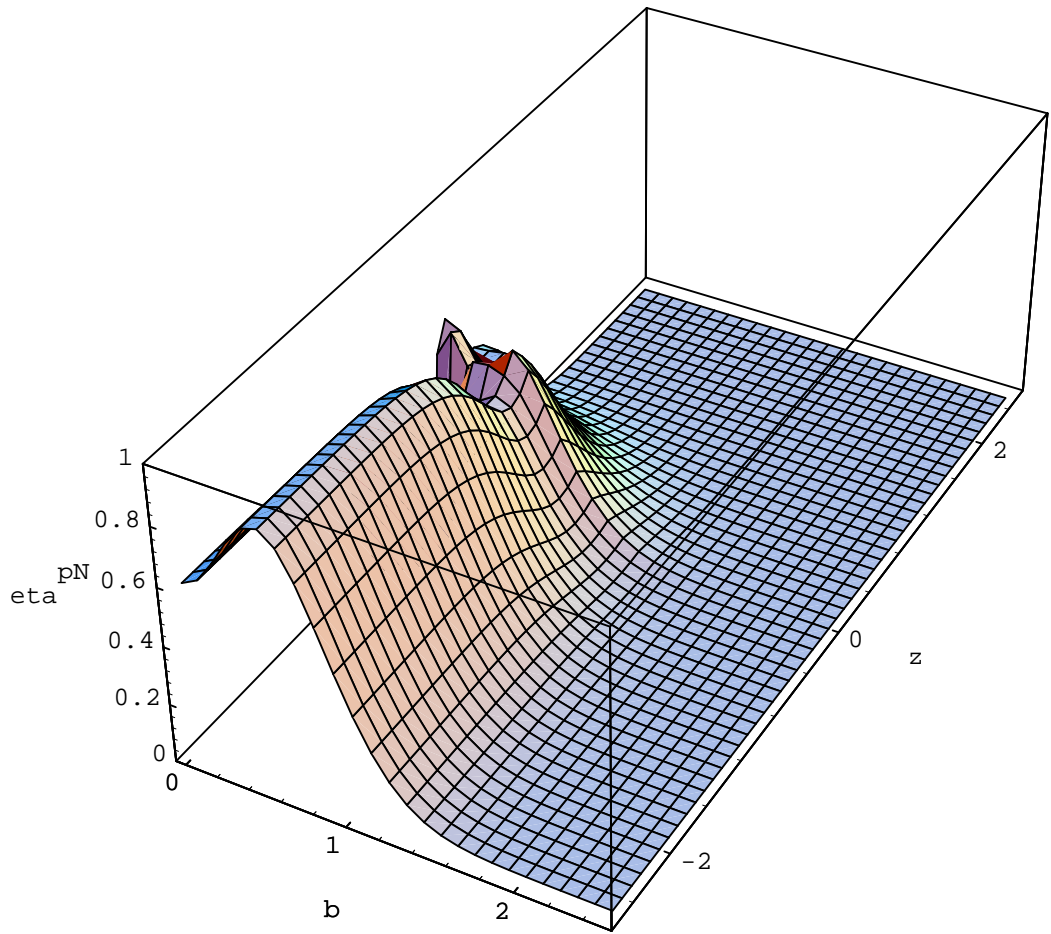

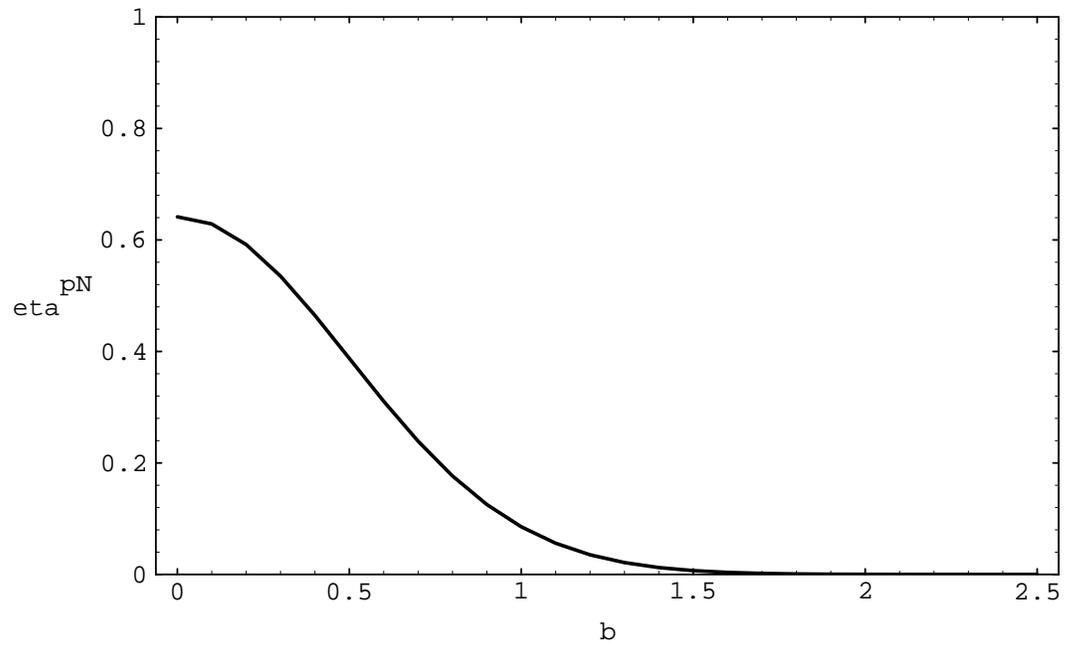

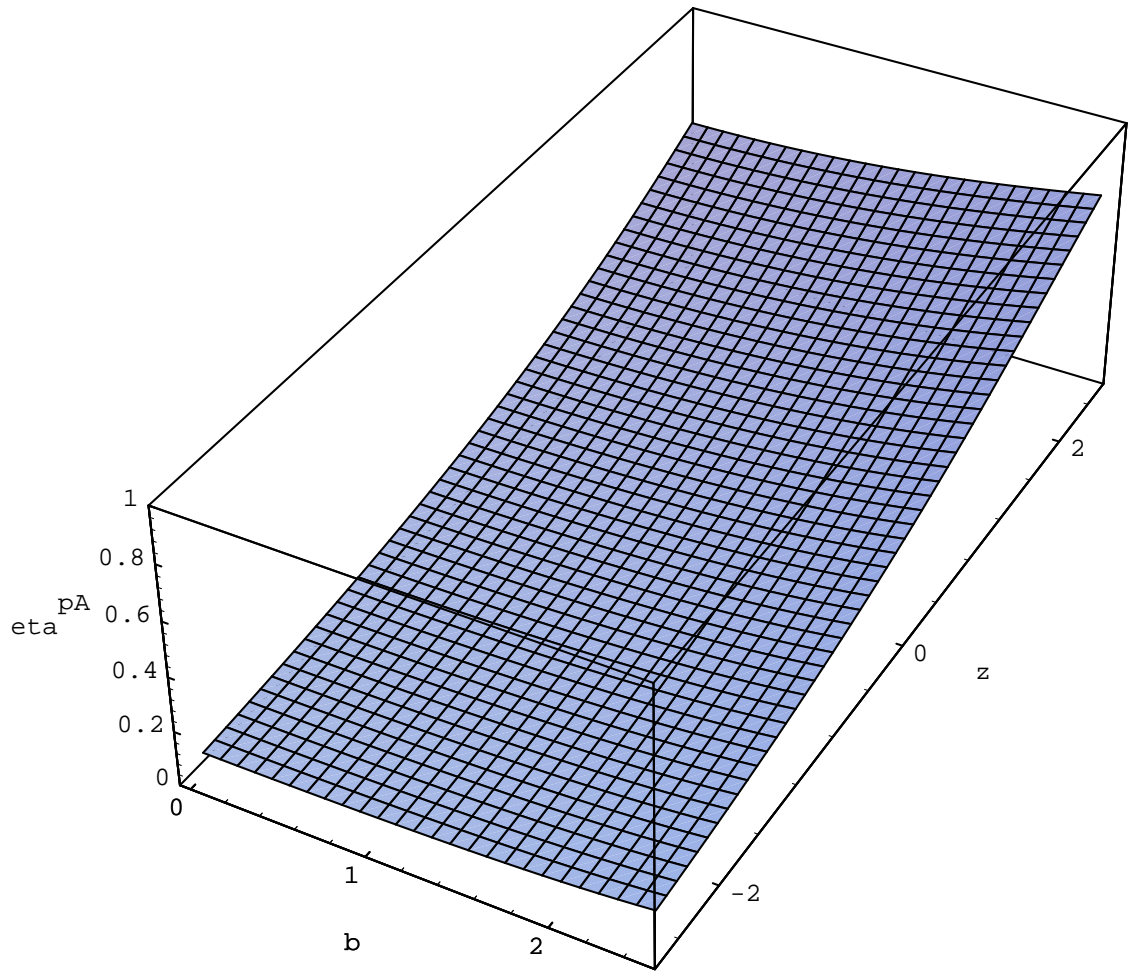

# Some remarks on off-shell scattering in the eikonal approximation

A.S. Rinat and M.F. Taragin

*Department of Particle Physics, Weizmann Institute of Science, Rehovot, 76100, Israel*

(March 27, 1996)

## Abstract

Using the Abel inversion for the eikonal phase as function of the interaction we derive simple integral relations between half and on-shell eikonal phases. A frequently used short-range approximation for the half off-shell phase and profile appears supported by the above-mentioned relation. We work out some examples and also address the half off-shell eikonal phase pertinent to a diffractive amplitude. The latter is relevant for a calculation of selected transparencies $\mathcal{T}$ of nuclei for a proton, knocked-out in selected semi-inclusive (SI) $A(e, e'p)X$ reactions. Some numerical results for $\mathcal{T}$ are given.
25.30 Fj, 24.10Eq

# I. INTRODUCTION.

The scattering of a projectile from an isolated scatterer is described by a $t$-matrix with arguments initial and final momenta $\boldsymbol{k}, \boldsymbol{k}'$, $|\boldsymbol{k}'| = |\boldsymbol{k}|$ and energy $E_k = k^2/2m$. A different situation prevails if a scatterer is embedded in a dense medium for which the average distance between scatterers $d$ is smaller than the interaction range $r_0$. A collision of the projectile with another particle may then take place before the projectile has left the range of the interaction of the first scatterer, i.e. before the system has time to conserve energy: the scattering occurs off-shell. In the momentum representation, scattering off the energy-shell corresponds to a situation where the available energy does not correspond to either the in-going or the out-going relative momenta $\boldsymbol{p}, \boldsymbol{p}'$.

Within a non-relativistic (NR) framework the most general off-shell $t$-matrix is computed from the Lippmann-Schwinger equation, which is driven by the (partial wave component of the) interaction $V$. For spinless particles it reads

$$\langle p'|t_l(E_k)|p\rangle = \langle p'|V_l|p\rangle + \int dp" p"^2 \langle p'|V_l|p"\rangle G_0(k,p")\langle p"|t_l(E_k)|p\rangle, \qquad (1)$$

with $G_0$ the free propagator. The half-off shell (HOS) and on-shell (ON) $t$-matrices $\tilde{t} = t^{HOS}$, $t^{ON} = t$ are obtained by the substitution $p \to k$ (or $p' \to k$), respectively both, in the totally off-shell (TOS) $t$-matrix. Integral equations have been constructed for their ratio [1] and difference [2], but there exist no *simple* dynamical relations between $\tilde{t}$ and $t^{ON}$.

The above equation or its variants, are the standard tools for the construction of the off-shell $t$-matrix as for instance occurring in the determination of the ground state and energy of any many-body, and in particular of a 3-body system, or in the calculation of the scattering amplitude of a projectile from a medium.

Below we shall consider those $t$-matrices in the description of high-energy semi- and totally inclusive scattering of electrons on composite targets, specifically $A(e,e'p)X$, $A(e,e')X'$. In these reactions the incident electron transfers to a struck nucleon a momentum $\boldsymbol{q} \gg \sqrt{\langle p^2 \rangle}$ with $p$, a typical momentum of a nucleon in the target in its ground state. We shall limit the discussion to the case where the nucleon with recoil momentum $|\boldsymbol{p}+\boldsymbol{q}| \approx q$ scatters from only one other nucleon, which has momentum $\boldsymbol{p}'$. Since $|\boldsymbol{p}'| \ll q$ the scattering occurs essentially under lab conditions with momentum $\approx q$. In view of its magnitude, a convenient description is provided by the spatial eikonal representation [3], where the struck nucleon enters and exits the scattering region with given impact parameter $\boldsymbol{b}$. The end points of the traversed longitudinal segment $(z, z')$ for those processes lie, at least in part, inside the interaction region which causes the scattering to be off-shell.

In a high-energy, NR description, using an intermediate interaction $V$, the accumulated off-shell phase and associated profile read

$$\chi_q^{TOS}(b, z, z') = -(1/v_q) \int_z^{z'} d\zeta V(b, \zeta) \tag{2a}$$

$$\Gamma_q^{TOS}(b, z, z') \equiv e^{i\chi_q^{TOS}(b,z,z')} - 1 \tag{2b}$$

On-shell scattering in the coordinate representation corresponds to the projectile which originates from, and exits into the interaction-free region. Denoting ON, HOS, TOS quantities by $\chi, \tilde{\chi}, \chi^{TOS}$, etc. one has

$$\Gamma_q(b) = \lim_{z \to -\infty, z' \to \infty} \Gamma_q^{TOS}(b, z, z') = e^{i\chi_q(b)} - 1 \tag{3a}$$

$$\chi_q(b) = -(1/v_q) \int_{-\infty}^{\infty} d\zeta V(\boldsymbol{b}, \zeta) \tag{3b}$$

We now elaborate on relations between half and on shell quantities in an eikonal representation.

## II. HALF AND COMPLETELY OFF-SHELL PHASES AND PROFILES.

Consider the following HOS phase and profile

$$\tilde{\chi}_q(b, z) = -(1/v_q) \int_z^{\infty} d\zeta V(\boldsymbol{b}, \zeta) \tag{4a}$$

$$\tilde{\Gamma}_q(b, z) \equiv e^{i\tilde{\chi}_q(b,z)} - 1 \tag{4b}$$

A construction of the desired HOF functions in an eikonal description could in principle start with the eikonalized version of the Lippmann-Schwinger equation (1) for $\tilde{t}$ (or equivalently for the full Greens function $G$) [5]. The corresponding profile $\tilde{\Gamma}_k(b,z)$ is then the one-dimensional Fourier transform $\tilde{t}_l(k,p)$, with $k$ the relative, instead of the lab momentum $q$ and $l+1/2 \to bk$.

Different approaches circumvent the solution of the above mentioned equation and address directly HOS quantities in the coordinate representation. A simple approximation holds for instance for short range-interactions, when one finds from Eq. (4)

$$\tilde{\chi}_q(b,z) \approx \theta(-z)\chi_q(b)$$
$$\tilde{\Gamma}_q(b,z) \approx \theta(-z)\Gamma_q(b) \tag{5}$$

We now turn to exact relations. Already in his first paper on eikonal scattering Glauber remarked that for a spherically symmetric, energy-independent $V$, Eq. (3a) can be inverted [3]. With $r = \sqrt{b^2 + z^2}$

$$V_q(r) = \frac{v_q}{\pi r}\frac{d}{dr}\int_r^\infty \frac{db' b' \chi_q(b')}{\sqrt{b'^2 - r^2}} = \frac{v_q}{\pi}\int_r^\infty \frac{db'[d\chi_q(b')/db']}{\sqrt{b'^2 - r^2}} \tag{6}$$

Harrington later remarked that Eqs. (2a) and (4a), in conjunction with (6) establish *in principle* a relation between eikonal forms of $t, \tilde{t}, t^{TOS}$, which however he did not elaborate [5]. Only recently in a study of the transparency of nuclei for protons knocked-out in a $A(e,e'p)X$ reaction, did Seki *et al* use the Abel inversion (6) in order to calculate $V_{NN}$ from scalar elastic $NN$ amplitudes [6]. $\tilde{\chi}$ was then computed by means of (4).

No one appeared drawn to a general exploration of an actual elimination of $V$ between (2a), (3a) and (6). In fact, this is a very simple affair with some noteworthy implications. One easily shows

$$\tilde{\chi}_q(b,z) = \theta(-z)\chi(b) + \begin{cases} -\frac{1}{\pi}\frac{z}{|z|}\int_r^\infty db'\frac{d\chi_q(b')}{db'}\cos^{-1}\left[\frac{|z|}{\sqrt{b'^2 - b^2}}\right] \\ \frac{1}{\pi}\frac{z}{|z|}\int_1^\infty \frac{d\beta}{\beta}\frac{\chi_q(\sqrt{b^2 + (\beta z)^2})}{\sqrt{\beta^2 - 1}} \\ \frac{1}{\pi}\frac{z}{|z|}\int_1^\infty \frac{d\beta}{\beta}\frac{\chi_q(\sqrt{(b/r_0)^2 + (\beta z/r_0)^2})}{\sqrt{\beta^2 - 1}}, \end{cases} \tag{7}$$

where in the third line above we introduced a scale parameter $r_0$, convenient for studying the small range behaviour. As a comparison with (5) shows, (7) expresses the HOS phase as the 0-range part and a remainder for a finite, not-necessarily small range.

Eq. (7) also makes explicit the intimate relation between the 0-range and the on-shell limit $z \to -\infty$. One checks

$$\lim_{r_0 \to 0} \tilde{\chi}_q(b, z) = \theta(-z)\chi(b), \tag{8a}$$

$$\lim_{z \to \infty} \tilde{\chi}_q(b, z) = 2\theta(-z) \lim_{z \to 0} \tilde{\chi}_q(b, z) = \theta(-z)\chi(b) \tag{8b}$$

where we used that the on-shell limit is twice the result for $z \to 0$.

The generalization of the above to TOS phases is trivial, once it is realized from (2) that $\chi, \tilde{\chi}, \chi^{TOS}$ are linearly related, thus

$$\chi_q^{TOS}(b, z, z') = -\frac{1}{v_q} \int_z^{z'} d\zeta V(b, \zeta) = \tilde{\chi}_q(b, z) - \tilde{\chi}_q(b, z') \tag{9}$$

The above expressions, relating the HOS and ON phases are actually surprising, when comparing the relations implied by (1) on the one hand, and (4), (7) and (9) on the other hand. Simple integrals instead of a solution of integral equations, provide the relation between off-and on-shell quantities. The fact that those are phases and not $t$-matrices or their Fourier transforms, the profiles, hardly matters in practice.

Eqs. (7) have been derived using non-relativistic dynamics, but the result does not contain the intermediate interaction $V$. As is not uncommon in eikonal theories, one is then led to postulate that results like (7) and (8) hold in a more general context, irrespective of the intermediate potential model. [1] In particular one would like to use those in the high-energy regime, where the $q$-dependence of the generally complex phase is more complicated than through $1/v_q$.

In the following Section we discuss a few examples.

---

[1] A relativistic generalization for the appropriate off-mass shell amplitude parallel to (1), is in principle provided by the Bethe-Salpeter equation for 2-particle scattering.

## III. EXAMPLES.

I) Gaussian interaction $V_G = \lambda e^{-r^2/r_0^2}$:

One readily checks

$$\chi_q(b) = -\frac{\sqrt{\pi} r_0 \lambda}{v_q} e^{-(b^2/r_0^2)} \tag{10a}$$

$$\tilde{\chi}_q(b,z) = \frac{1}{2}\text{Erfc}(z/r_0)\chi(b) \tag{10b}$$

$$\lim_{r_0 \to 0} \tilde{\chi}_q(b,z) = \chi(b)\left[\theta(-z)\sqrt{\pi} + [\theta(z) - \theta(-z)]\frac{r_0}{\sqrt{\pi}|z|}e^{-(z/r_0)^2}\right], \tag{10c}$$

For $r_0 < z$, (i.e. for distances $z$ between the scattering centers larger than the range $r_0$, (10c) provides exponentially decreasing corrections to the 0-range result (8). Similar results may be expected for interactions bounded by a Gaussian.

II) Lorentzian interaction $V_L(r) = (\lambda/\pi)\left[\gamma/(z^2 + B^2)\right]$ with $B \equiv \sqrt{b^2 + \gamma^2/4}$:

$$\chi_q(b) = -\frac{\lambda\gamma}{v_q B} \tag{11a}$$

$$\tilde{\chi}_q(b,z) = \chi_q(b)\pi^{-1}\sin^{-1}\left[\frac{B}{\sqrt{B^2 + z^2}}\right] \tag{11b}$$

III) Diffractive scattering:

In the general discussion in Section II, as well as in the two mentioned examples we did not juxtapose a HOS $t$-matrix with its corresponding Fourier transform, the HOS profile $\tilde{\Gamma}$, but instead with $\tilde{\chi}$. In a last example we start from an on-shell profile corresponding to an amplitude which, for high energies describes diffractive scattering in the forward angular cone. A simple parameterization in the impact parameter representation reads

$$\Gamma_q(b) \approx -\frac{\sigma^{tot}}{2}(1 - i\tau)A_q(b, Q_0) \tag{12a}$$

$$A_q(b, Q_0) = \left(Q_0^2(q)/4\pi\right)\exp[-(bQ_0(q)/2)^2], \tag{12b}$$

$$\chi_q(b) = -i\ln[\Gamma_q(b) + 1] \tag{12c}$$

with $\tau$ the ratio of the real to imaginary part of the forward elastic amplitude and $A_q(b, Q_0)$ a function, representing the dependence on a range $r_0 \approx 2/Q_0(q)$.

In a numerical example we take $pN$ scattering at momentum $q=4.49$ GeV, the value chosen by Seki et al [6]. For it, the averaged $pp$ and $pn$ parameters are $Q_0=2.8$ fm$^{-1}$, $\sigma^{tot}=43.2$ mb, $\tau \approx -0.41$. The corresponding differential elastic cross section $d\sigma(Q^2)/d\Omega \approx (k\sigma^{tot}/2\pi)^2(1+\tau^2)\exp[-(2Q^2/Q_0)^2]$, with $Q$ the momentum transfer, accounts for the forward angle data over more than two decades.

Although not directly related to our topic, we have used (6) for a computation of the effective, central $NN$ interaction $V_{NN}$, which produces the diffractive phase (12c). In the range $1.2 \leq q(\text{GeV}) \leq 4.5$ investigated there is a $q$-dependence, for instance in $\text{Re}V(0), \text{Im}V(0)$, reflecting the same in the above phase. We further mention an estimated range of $\text{Re}V$ of $\approx 0.6$ fm, to be compared with a range parameter from the *amplitude* $r_{A,0} \approx 2/Q_0 \approx 0.72$ fm.

Next we consider the HOS absorption function $\tilde{\eta}^{pN}(\bm{r}) = 1 - e^{-2\text{Im}\tilde{\chi}(\bm{r})}$, which is non-vanishing even on the $NN$ level if inelastic channels are open. We shall compare the following expressions

$$\tilde{\eta}_q^{pN}(\bm{r}) = 1 - e^{-2\text{Im}\tilde{\chi}_q(\bm{r})} \tag{13a}$$

$$\approx \theta(-z)[1 - e^{-2\text{Im}\chi_q(b)}] \approx \theta(-z)[\sigma_q^{tot} A_q(b, Q_0) - \sigma_q^{tot,el} A_q(b, Q_0/\sqrt{2})] \tag{13b}$$

$$\approx \theta(-z)\sigma_q^{tot,inel} \tag{13c}$$

Eqs. (13a) uses the HOS phase from (7), while (13b) exploits the short range limit (8a) and the parameterization (12). Finally, Eq. (13c) is the 0-range limit of the latter ($A(b, Q_0) \to \delta^2(\bm{b})$).

In virtually all treatments of semi-inclusive processes thus far one relied on the short range limits (13b) or (13c) instead of the proper eikonal expression (13a) [9,10,13]. In spite of their formal similarity, the explicit forms differ manifestly. Eq. (13b) makes direct use of the on-shell $\chi$ and is, for $z > 0$ or $< 0$, independent of $z$. In contradistinction, Eq. (13a) without a restriction on the range, employs the on-shell phase $\chi$ in (7a) in order to construct a $z$-dependent HOS phase $\tilde{\chi}$ and $\tilde{\eta}^{pN}$.

Fig. 1a is a 3-dimensional plot of $\tilde{\eta}^{pN}(|\bm{b}|, z)$ for $q=4.5$ GeV and one clearly discerns short-range characteristics. Regrettably we have no details regarding the $NN$ amplitudes used by Seki et al, which prevents a comparison with their result [6]. Fig. 1b shows the small-range limit (13b) for any $z < 0$.

This concludes our remarks on off-shell quantities per se. In Section IV we embed those results in selected high-energy, semi-inclusive scattering.

## IV. OFF-SHELL PHASES IN INCLUSIVE SCATTERING.

In a high-energy, $A(e, e'p)X$ process a proton is knocked-out somewhere inside the target, propagates and is ultimately detected. A description of a collision with a second nucleon on its way out, requires per definition a half off-shell quantity, in our case $\tilde{\chi}$ (see for instance [4]). A characteristic medium property for that reaction is the nuclear transparency $\mathcal{T}$ for the knocked-out proton, defined as the ratio of the experimental yield and a hypothetical one with no interaction between that proton and the core.

The above transparencies depend on the experimental conditions of the experiment, in particular on those of the knocked-out proton [9,10]. Particularly simple expressions result if the above yields are integrated over the energy loss of the electron, over the proton momentum or both. In relation to recent work by Seki et al [6] we discuss here the special case $\mathcal{T} \to \mathcal{T}^{PE}$. For it, we limit ourselves to FSI, due to binary collisions between the knocked-out proton and core nucleons. In the first cumulant approximation, one obtains for the above transparency [4,6,10,11]

$$\mathcal{T}_q^{PE} = \int d\boldsymbol{r}_1 \rho(\boldsymbol{r}_1) \exp[\eta_q^{pA}(\boldsymbol{r}_1)] \tag{14a}$$

$$\eta_q^{pA}(\boldsymbol{r}_1) = -(A-1) \int d\boldsymbol{r}_2 \frac{\rho_2(\boldsymbol{r}_1, \boldsymbol{r}_2)}{\rho(\boldsymbol{r}_1)} \tilde{\eta}^{pN}(\boldsymbol{r}_1 - \boldsymbol{r}_2) \tag{14b}$$

$$\approx -(A-1) \int d\boldsymbol{r}' \rho(\boldsymbol{r}_1 - \boldsymbol{r}') g(\boldsymbol{r}') \tilde{\eta}_q^{pN}(\boldsymbol{r}'), \tag{14c}$$

where $\rho$ is the single-particle density and $g(\boldsymbol{r})$ the $NN$ pair-distribution function. The latter has been assumed not to depend on the center-of-mass of the two particles involved. Eq. (14c) establishes the relation between the local *nuclear* and *nucleonic* absorption functions

(13). [1]

We have computed the nuclear absorption function using the expressions (13) for $\tilde{\eta}^{pN}$, and found that the averaging over the density in (14b) largely removes the non-negligible differences in the latter (cf. Figs. 1a,b). In Fig. 2 we show $\eta^{pA}$ for Fe, using the more general $\eta^{pN}$ from (13a). In spite of the above smearing there survive characteristic differences for $z \leq 0$ and $z \geq 0$.

In Table I we entered transparencies from doubly integrated SI cross sections for a variety of targets. Pairs of columns correspond to results, computed with and without correlations, embodied by the pair correlation function $g(r)$. An equation number at the heading indicates the expression for the off-shell phase function, which has been used in the calculation of the $pA$ absorption function, Eq. (14).

We start with predictions including correlations. In spite of the differences spelled out after Eq. (13) there is surprisingly little difference between results from (13a) and (13b). The replacement of a finite range $r_0 \approx 0.7$fm $\to 0$ reduces $\mathcal{T}$ by a few %. Neglect of correlations increases the absorption and thus decreases $\mathcal{T}$.

The same relative changes have also been reported in [6], but our transparencies are sizably larger than those of Seki et al. Again we lack information which enables tracing of the differences. Both groups use (13b) in (14b) as did Nikolaev et al [9,10]. We recall that in going from (14b) to (14c) we introduced a pair correlation function and assumed it not to depend on the center-of-mass coordinate $R$ of a pair. In contradistinction, Seki et al mention in [6] a 2-particle density function, integrated over $R$, without stating apparent approximations involved in the replacement of $\rho_2(\boldsymbol{r}_1, \boldsymbol{r}_2)$ by it in (14b). Since the application to SI inclusive reactions $A(e, e'p)X$ is only a side issue in this note on off-shell eikonal phases, the above is not pursued any further here. For the same reason we do not add remarks on totally inclusive processes where the knocked-out proton is not detected and consequently requires a TOS phase for the description of its FSI [13].

---

[1] The above FSI factor $\exp[\eta^{pA}]$ is identical to $R$ in Ref. [9] for the appropriate experimental conditions.

We summarize our results for the two points worked out in this note. The first deals with half-off shell phases in an eikonal description. Seki et al recently constructed those from a given central interaction, in turn computed from the on-shell phase. Above we derived a direct integral relation between half off-shell and on-shell phases. From it, one finds a proper justification for the use of a simple standard short-range approximation as well as finite-range corrections to it.

The second point employs the above off-shell phase to compute special nuclear transparencies from integrated semi-inclusive $A(e, e'p)X$ cross sections. We tested the influence on $\mathcal{T}$ of various forms of $\tilde{\chi}^{pN}$, and for each, the role of $NN$ correlations in the target. With only small higher order FSI effects [6], we estimate that the remaining major source of uncertainty in computed results for the above, as well as for less restrictive transparencies may probably be resolved through an accurate inclusion of precisely calculated two-particle densities.

Table I

|  | Eq. (13a) | | Eq. (13b) | | Eq. (13c) | |
| --- | --- | --- | --- | --- | --- | --- |
| target | $g \neq 1$ | $g = 1$ | $g \neq 1$ | $g = 1$ | $g \neq 1$ | $g = 1$ |
| C | .713 | .683 | .710 | .665 | .726 | .638 |
| O | .672 | .643 | .668 | .623 | .684 | .597 |
| Al | .591 | .560 | .586 | .540 | .604 | .516 |
| Ca | .550 | .521 | .543 | .501 | .561 | .479 |
| Fe | .502 | .475 | .496 | .455 | .513 | .434 |
| Au | .338 | .315 | .332 | .299 | .347 | .284 |

Nuclear transparencies $\mathcal{T}_q^{PE}$ from semi-inclusive $A(e, e'p)$ cross sections, integrated over missing momentum and energies. Results are for the binary collision approximation (14a), using (13). Pairs of columns correspond to the included, respectively neglected $NN$ correlations.

**Figure Captions.**

Figs. 1a,b. The local $pN$ absorption function (13a) and its small-range limit (13b) for $q = 4.49$ GeV.

Figs. 2. The local $pA$ absorption coefficient $\eta^{pA}(b, z)$, Eq. (14) for Fe, computed with (13a) for $\eta^{pN}(b, z)$.